\documentclass[twocolumn,superscriptaddress,prx]{revtex4-1}
\usepackage{graphicx}
\usepackage{amsmath}
\usepackage{amsthm}
\usepackage{amssymb}
\usepackage{latexsym}
\usepackage{array}
\usepackage{hyperref}
\usepackage{float}
\usepackage{amsfonts}
\usepackage{mathrsfs}
\usepackage{verbatim}
\usepackage{bbold}
\usepackage[normalem]{ulem}

\usepackage{upgreek}

\usepackage{makecell}
\usepackage{adjustbox,lipsum}

\usepackage{algorithm}
\usepackage{algpseudocode}

\usepackage{color}

\newcommand{\bra}[1]{\left<#1\right|}
\newcommand{\ket}[1]{\left|#1\right>}
\newcommand{\abs}[1]{\left|#1\right|}

\newcommand{\expt}[1]{\left<#1\right>}
\newcommand{\braket}[2]{\left<{#1}|{#2}\right>}

\newcommand{\tr}[1]{\mbox{Tr}{#1}}

\newtheorem{proposition}{Proposition}
\newtheorem{lemma}{Lemma}

\begin{document}

\title{Quantum Amplitude-Amplification Eigensolver: A State-Learning-Assisted Approach beyond Energy-Gradient-Based Heuristics}

\author{Kyunghyun~Baek}
\affiliation{Institute for Convergence Research and Education in Advanced Technology, Yonsei University, Seoul 03722, Republic of Korea}%
\affiliation{Department of Quantum Information, Yonsei University, Incheon 21983, Republic of Korea}%

\author{Seungjin~Lee}
\affiliation{IMEC, Kapeldreef 75, Leuven, Belgium, 3001}%

\author{Joonsuk~Huh}
\affiliation{Department of Chemistry, Yonsei University, Seoul 03722, Republic of Korea}%
\affiliation{Department of Quantum Information, Yonsei University, Incheon 21983, Republic of Korea}%

\author{Dongkeun~Lee}
\affiliation{Center for Quantum Information R\&D, Korea Institute of Science and Technology Information, Daejeon 34141, Republic of Korea}%

\author{Jinhyoung~Lee}
\affiliation{Department of Physics, Hanyang University, Seoul, 04763, Republic of Korea}%

\author{M.~S.~Kim}
\affiliation{Blackett Laboratory, Imperial College London, London SW7 2AZ, United Kingdom}%

\author{Jeongho~Bang}\email{jbang@yonsei.ac.kr}
\affiliation{Institute for Convergence Research and Education in Advanced Technology, Yonsei University, Seoul 03722, Republic of Korea}%
\affiliation{Department of Quantum Information, Yonsei University, Incheon 21983, Republic of Korea}%

\date{\today}

\date{\today}

\begin{abstract}
Ground-state estimation lies at the heart of a broad range of quantum simulations. Most near-term approaches are cast as variational energy minimization and thus inherit the challenges of problem-specific energy landscapes. We develop the quantum amplitude-amplification eigensolver (QAAE), which departs from the variational paradigm and instead coherently drives a trial state toward the ground state via quantum amplitude amplification. Each amplitude-amplification round interleaves a reflection about the learned trial state with a controlled short-time evolution under a normalized Hamiltonian; an ancilla readout yields an amplitude-amplified pure target state that a state-learning step then re-encodes into an ansatz circuit for the next round---without evaluating the energy gradients. Under standard assumptions (normalized $\hat{H}$, a nondegenerate ground-state, and a learning update), the ground-state overlap increases monotonically per round and the procedure converges; here, a per-round depth bound in terms of the ansatz depth and Hamiltonian-simulation cost establishes hardware compatibility. Cloud experiments on IBMQ processor verify our amplification mechanism on a two-level Hamiltonian and a two-qubit Ising model, and numerical benchmarks on $\mathrm{H}_2$, $\mathrm{LiH}$, and a $10$-qubit longitudinal-and-transverse-field Ising model show that QAAE integrates with chemistry-inspired and hardware-efficient circuits and can surpass gradient-based VQE in accuracy and stability. These results position QAAE as a variational-free and hardware-compatible route to ground-state estimation for near-term quantum simulation.
\end{abstract}

\maketitle

\section{Introduction}\label{Sec:1}

Estimating the ground state of an interacting many‑body Hamiltonian is central to quantum simulation~\cite{Georgescu2014}. However, accurate ground‑state determination is formidable: the Hilbert space grows exponentially with the number of orbitals (or qubits), and the local Hamiltonian problem is QMA‑hard~\cite{Kempe2006}. Classical modern numerics beyond exact diagonalization exploit structures but face intrinsic limits. For example, Tensor‑network techniques (e.g., DMRG/MPS, TEBD) perform well for area‑law states, but the required bond dimension swells with entanglement and/or long‑range couplings, rapidly increasing the cost~\cite{Schollwock2005,Orus2014,Vidal2004}. Stochastic schemes (e.g., quantum Monte-Carlo) are efficient only in sign‑problem‑free cases~\footnote{Otherwise the average sign decays---often exponentially in system size or inverse temperature---sending variances skyrocketing~\cite{Troyer2005}.}. Thus, for strongly correlated and/or higher‑dimensional systems, these tools often lose accuracy or require prohibitive resources.

Meanwhile, quantum algorithms promise advantages. Chief among them is quantum phase estimation (QPE), which---given a trial state with nonzero ground‑state overlap---projects onto an eigenstate and estimates its energy with Heisenberg‑limited precision~\cite{Abrams1999,Aspuru2005,Higgins2007,Lin2022}. However, QPE‑based eigensolvers require an $m$‑qubit phase register for $\mathcal{O}(2^{-m})$ resolution, long sequences of multi-controlled-unitary evolutions, and an inverse quantum Fourier transform (or semi‑classical variants) in practice. For generic many‑body Hamiltonians, this QPE circuit incurs accuracy‑dependent overhead~\cite{Low2017,Childs2021}; the resulting depth and fidelity typically exceed the near‑term capabilities. Iterative or Bayesian QPE can reduce ancillas, but still demands long circuit depths, and the success probability hinges on the initial overlap---often prompting amplitude amplification or tailored preparation, further increasing the implementation depth~\cite{Griffiths1996,Fomichev2024,Lee2025}. Thus, QPE remains the asymptotic gold standard, but its practical realization is expected to await the fault‑tolerant architectures with robust quantum error correction, motivating hardware‑conscious alternatives in the interim.

Very recently, more resource‑efficient and near‑term realizable approaches have emerged as candidates for tangible quantum advantage~\cite{Cerezo2021a,Bharti2022}. Foremost among them is the variational quantum eigensolver (VQE)~\cite{Peruzzo2014,Kandala2017,Tilly2022}, which leverages the variational principle $\langle\Psi|\hat{H}|\Psi\rangle\ge\lambda_0$ to recast ground‑state estimation~\cite{Mcclean2016,Yuan2019}:
\begin{eqnarray}
\min_{\theta}{E(\theta)} := \bra{\Psi(\boldsymbol\theta)}\hat{H}\ket{\Psi(\boldsymbol\theta)},
\label{eq:vqe-objective}
\end{eqnarray}
where $\lambda_0$ is the ground‑state energy of a given Hamiltonian $\hat{H}$ and the trial state $\ket{\Psi(\boldsymbol\theta)}$ is prepared by a parametric quantum circuit, often-called ansatz; the parameter vector $\boldsymbol\theta$ is updated by a classical optimizer. In practice, $\hat{H}$ is decomposed as $\hat{H}=\sum_{l=1}^{L}w_l \hat{P}_l$ (e.g., Pauli strings or few‑body terms), the expectation $E(\theta)$ is estimated from repeated measurements, and gradients are obtained via parameter‑shift or finite‑difference rules, yielding a classical-quantum hybrid interplay~\footnote{In VQE, the quantum processor supplies (noisy) unbiased estimates of $E(\theta)$ and its gradients, while a classical routine (e.g., SPSA/Adam/COBYLA) updates $\theta$; see Ref.~\cite{Lavrijsen2020} for surveys of hybrid workflows.}.

Despite many proof‑of‑concept successes, the VQE framework faces obstacles. (i) Local minima: Finite‑shot noise and hardware imperfections yield nonconvex objectives, making convergence sensitive to initialization~\cite{Nakanishi2020,Wierichs2020}. (ii) Barren plateaus: For sufficiently expressive or randomly initialized ansatz, the energy-gradient variance can vanish exponentially with system size~\cite{McClean2018,Cerezo2021b,Mele2022,Bittel2021}. The noises can further induce plateauing~\cite{Wang2021}. (iii) Expressivity-vs-trainability trade‑off: Chemistry‑inspired ansatz (e.g., UCCSD/ADAPT-type) mitigates plateaus but incurs large parameter counts; hardware‑efficient circuits are shallow, yet often harder to train~\cite{Sim2019,Holmes2022}. (iv) Measurement overhead: Shot complexity scales with estimator variance and the number of Hamiltonian terms~\cite{Wecker2015,Crawford2021}. Collectively, these factors limit the reliability of the variational methods and motivate alternative algorithms that avoid explicit energy‑landscape search based on energy-gradient, while offering convergence guarantees under hardware‑aware resource bounds.

In this work, we develop the quantum amplitude-amplification eigensolver (QAAE), that coherently steers a trial state toward the ground state. Each round applies a reflection about the learned trial state and a short‑time controlled evolution under a normalized Hamiltonian; an ancilla readout yields an amplitude‑amplified pure state that is re‑encoded into an ansatz circuit for the next round---\emph{without} evaluating the energy gradients. This amplify--learn loop increases the ground‑state overlap while avoiding the exploration of rugged Hamiltonian landscapes and can mitigate gradient‑pathologies typical of variational heuristics. We validate this amplification mechanism on IBMQ processor (a two‑level Hamiltonian and a two‑qubit Ising model), and benchmark QAAE on $\mathrm{H}_2$ ($4$ qubits), $\mathrm{LiH}$ ($12$ qubits), and longitudinal‑and‑transverse‑field Ising model ($10$ qubits), where the chemistry‑inspired and hardware‑efficient PQCs integrate naturally. In particular, the benchmark on $10$‑qubit Ising model attains higher accuracy and reliability than VQE under matched resources. 

\section{Quantum Amplitude Amplification Eigensolver}\label{Sec:2}

\begin{figure*}[t]
\center
 \includegraphics[width=0.62\textwidth]{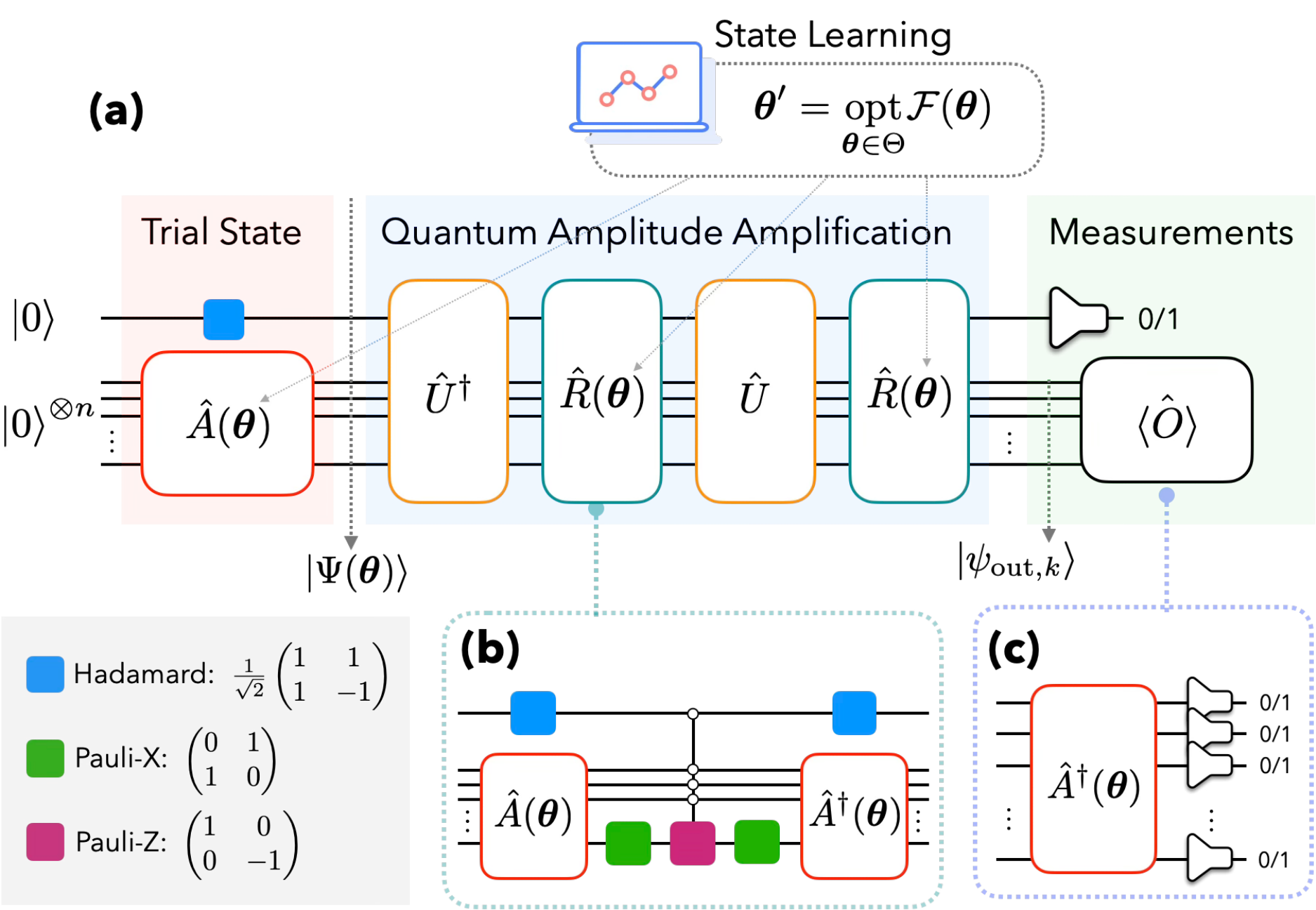}
\caption{Schematic of our QAAE setup. (a) The circuits for generating the trial state $\ket{\Psi(\boldsymbol\theta)}$ and implementing the ground-state amplitude amplification $\hat{T}(\boldsymbol\theta)$. (b) The reflecting operation $\hat{R}(\boldsymbol\theta)$ is implemented by using $\hat{A}(\boldsymbol\theta)$ and some fundamental gates (Hadamard, Pauli-X($\hat{\sigma}_x$), Pauli-Z($\hat{\sigma}_z$), and $q$-Toffoli, etc.). (c) The state-learning [as in Eq.~(\ref{eq:opt})] is performed for $\ket{\varphi_{\text{out},k}}$. Then, a new vector $\boldsymbol\theta'$ is identified and the circuit elements $\hat{A}(\boldsymbol\theta)$ and $\hat{R}(\boldsymbol\theta)$ are (re)dialed: $\boldsymbol\theta \to \boldsymbol\theta'$.}
\label{fig:schematic}
\end{figure*}

To implement QAAE, we start by preparing a trial state, denoted as $\ket{\Psi(\boldsymbol\theta)}$. Here, $\boldsymbol\theta$ is a vector consisting of the adjustable parameters $\theta_j$ ($j=0,1,\ldots,K-1$): i.e., $\boldsymbol\theta=(\theta_0, \theta_1, \ldots, \theta_{K-1})^T$. The trial-stat encoding is performed by using an ansatz, denoted as $\hat{A}(\boldsymbol\theta)$, such that
\begin{eqnarray}
\ket{\Psi(\boldsymbol\theta)} = \ket{+} \otimes \ket{\alpha(\boldsymbol\theta)} = \ket{+} \otimes \hat{A}(\boldsymbol\theta)\ket{\mathbb{0}},
\label{eq:ansatz}
\end{eqnarray}
where $\ket{\mathbb{0}}=\ket{0}^{\otimes q}$ and $q$ is the number of qubits to model the given ($2^q$-dimensional) Hamiltonian $\hat{H}$. Here, the ansatz-estimated ground state $\ket{\alpha(\boldsymbol\theta)} = \hat{A}(\boldsymbol\theta)\ket{\mathbb{0}}$ is expected to evolve towards the true ground state. The single-qubit ancilla state $\ket{+} = \frac{1}{\sqrt{2}}\left(\ket{0} + \ket{1}\right)$ is used for the stability of the process~\cite{Bang2015}. Then, we introduce a unitary process associated with a normalized Hamiltonian $\hat{H}$ (i.e., the eigenvalues are in $(0,1]$), defined as
\begin{eqnarray}
\hat{U}= \sum_{k=0,1} i^k \ket{k}\bra{k} \otimes e^{(-1)^k i \omega \hat{H}},
\label{eq:U}
\end{eqnarray}
which is supposed to be designed efficiently~\cite{Lloyd1996,Childs2021}. Here, $\omega$ is set to be $\tfrac{\pi}{4}$. With this unitary $\hat{U}$, we can define an operation for a single round of the amplification as
\begin{eqnarray}\label{eq:T}
\hat{T}(\boldsymbol\theta) = \hat{R}(\boldsymbol\theta) \hat{U} \hat{R}(\boldsymbol\theta) \hat{U}^\dagger,
\end{eqnarray}
where $\hat{R}(\boldsymbol\theta)=\hat{\openone} - 2 \ket{\Psi(\boldsymbol\theta)}\bra{\Psi(\boldsymbol\theta)}$, often-called Householder reflection~\cite{Brassard2002,Ambainis2004,Ivanov2006}. Given a set of the parameters $\boldsymbol\theta$, $\hat{R}(\boldsymbol\theta)$ can be implemented by using $\hat{A}(\boldsymbol\theta)$ and $\hat{A}^\dagger(\boldsymbol\theta)$. Schematic of the setup is depicted in Fig.~\ref{fig:schematic}. 

{\em Amplify--Learn loop.}---We describe how QAAE process runs. The process comprises the two essential steps:

\medskip
({\bf A.1}) {\bf Amplify}: $\hat{T}(\boldsymbol\theta)$ is applied to $\ket{\Psi(\boldsymbol\theta)}$ for chosen $\boldsymbol\theta$. Subsequently, by performing the projection measurement $\hat{P}_k = \ket{k}\bra{k}$ ($k \in \{ 0,1 \}$) on the ancillary qubit, we have the output state as
\begin{eqnarray}
\ket{\varphi_{\text{out},k}} = \frac{\bra{k} \hat{T}(\boldsymbol\theta) \ket{\Psi(\boldsymbol\theta)}}{\sqrt{p_k}},
\label{eq:output_state}
\end{eqnarray}
where $p_k$ is the probability of measuring $\ket{k}$, with the calculated value of $\frac{1}{2}$. The state $\ket{\varphi_{\text{out},k}}$ becomes closer to the ground state compared to $\ket{\alpha(\boldsymbol\theta)}$ (as shown later).

\medskip
({\bf A.2}) {\bf Learn}: The trial state and $\hat{T}(\boldsymbol\theta)$ are reconfigured by the parameter updates: $\boldsymbol\theta \to \boldsymbol\theta'$. The main idea is to use $\ket{\varphi_{\text{out},k}}$ in the new trial state, enabling the step ({\bf A.1}) to reproduce an output closer to the ground state. To realize this idea, we prepare a next-round trial state
\begin{eqnarray}
\ket{\alpha(\boldsymbol\theta')} = \hat{A}(\boldsymbol\theta')\ket{\mathbb{0}} = \ket{\varphi_{\text{out},k}}
\label{eq:A_new}
\end{eqnarray}
with the updated $\boldsymbol\theta'$; thus, we have to learn $\ket{\varphi_{\text{out},k}}$. We formulate the task of learning $\ket{\varphi_{\text{out},k}}$ as:
\begin{eqnarray}
\underset{\boldsymbol\theta}{\mathrm{opt}} \mathcal{F}(\boldsymbol\theta),~\text{subject to}~\boldsymbol\theta \in \Theta,
\label{eq:opt}
\end{eqnarray}
where $\text{opt}_{\boldsymbol\theta} \in \{ {\arg\max}_{\boldsymbol\theta}, {\arg\min}_{\boldsymbol\theta} \}$, $\mathcal{F}(\boldsymbol\theta)$ is an objective function, and $\Theta$ represents a parameter space. The size and landscape of $\Theta$ depends on the underlying structure of the learning $\hat{A}(\boldsymbol\theta)$~\cite{Zhao2024}. Crucially, the learning target is a fixed pure state rather than an energy functional; hence, no energy-gradient‑based heuristics are employed. 

\medskip
An iteration of ({\bf A.1})--({\bf A.2}) yields a sequence whose ground‑state overlap increases round by round. A practical halting rule compares the successive‑round energies or output-state overlaps. If the process does not halted within a maximum round/shot budget, it returns ``task failure.'' The detailed line-by-line algorithmic description of our QAAE process is presented in {\bf Algorithm~\ref{alg:QAAE}}.
\begin{algorithm}[H]
\caption{Our QAAE process}\label{alg:QAAE}
\begin{algorithmic}[1]
\Require{$\ket{\mathbb{0}}$, $\ket{+}$, $\hat{A}(\boldsymbol\theta)$, $\hat{A}^\dagger(\boldsymbol\theta)$, $\hat{U}$, $\hat{U}^\dagger$, $\hat{R}_0$, $\hat{P}_k$}
\State Task: Given Hamiltonian $\hat{H}$, find its ground state $\ket{\lambda_0}$
\State Initialize: 
	\State $\ket{{\mathbb{0}}}=\ket{0}^{\otimes q}$, $\hat{R}_0 = \hat{\openone} - 2\ket{\mathbb{0}}\bra{\mathbb{0}}$
	\State $iter=0$, $halt \gets \emptyset$, $\Xi \gets \text{an arbitrary large number}$
	\State $\boldsymbol\theta \gets \{ Rand(\theta_j) \}$	\Comment{$Rand(\cdot)$: Random function}
    \While{$halt = \emptyset$ AND $iter < \Xi$}
	\State $\ket{\Psi(\boldsymbol\theta)} \gets \ket{+} \otimes \hat{A}(\boldsymbol\theta)\ket{\mathbb{0}}$
	\State $\hat{T}(\boldsymbol\theta) \gets \hat{R}(\boldsymbol\theta) \hat{U} \hat{R}(\boldsymbol\theta) \hat{U}^\dagger$ where $\hat{R}(\boldsymbol\theta) = \hat{A}(\boldsymbol\theta)\hat{R}_0\hat{A}^\dagger(\boldsymbol\theta)$
        \State Apply $\hat{T}(\boldsymbol\theta)$ to $\ket{\Psi(\boldsymbol\theta)}$
        \State Measurement $\hat{P}_{k} = \ket{k}\bra{k}$ ($k=0,1$)
        \State Obtain the output state $\ket{\varphi_{\text{out},k}}$	\Comment{see Eq.~(\ref{eq:output_state})}
        \If {$k=0$ (or $k=1$)}
     	   		\State State learning: $\boldsymbol\theta' = {\mathrm{opt}_{\boldsymbol\theta \in \Theta}} \mathcal{F}(\boldsymbol\theta)$
        \EndIf
        \If {halting condition is met}
        		\State {$halt \gets True$} 
	\EndIf
        \State $\boldsymbol\theta \gets \boldsymbol\theta'$
        \State $iter \gets iter+1$
    \EndWhile
    \If {$iter \ge \Xi$}
	{\Return ``algorithm failure''}
    \EndIf
    \State \Return $\boldsymbol\theta$ and $\ket{\varphi_{\text{out},k}} \left( \simeq \ket{\lambda_0} \right)$ \Comment{Solution}
\end{algorithmic}
\end{algorithm}

\section{Theoretical Analysis}\label{Sec:3}

\subsection{Converging behavior} 

We analyze the amplify--learn loop by reminding the ({\bf A.1}) amplify and ({\bf A.2}) learn steps. Throughout, we assume a normalized Hamiltonian $\hat{H}$ with a nondegenerate ground state $0 < \lambda_0 < \lambda_1 \le \cdots \le \lambda_{2^q-1} \le 1$, the ancilla prepared in $\ket{+}$, and the controlled short-time evolution $\hat{U}$ as in Eq.~(\ref{eq:U}) with the evolution-time parameter $\omega = \tfrac{\pi}{4}$; reflections and one-round operator are as in Eqs.~(\ref{eq:ansatz})--(\ref{eq:T}).

{\em Amplify: one-step improvement.}---Let the system trial state be decomposed in the eigenbasis as
\begin{eqnarray}
\ket{\alpha(\boldsymbol\theta)}=\sum_{j=0}^{d-1}\gamma_j\ket{\lambda_j}, \quad \Gamma_{\ket{\psi}}:=\abs{\braket{\lambda_0}{\psi}}^2,
\label{eq:psi_eig}
\end{eqnarray}
where $d=2^q$. Define the transition amplitude:
\begin{eqnarray}
\mathcal{W} := \bra{+,\alpha(\boldsymbol\theta)} \hat U \ket{+,\alpha(\boldsymbol\theta)} = \abs{\mathcal{W}} e^{i\frac{\pi}{4}}.
\label{eq:W_def}
\end{eqnarray}
Here, $\abs{\mathcal{W}} = \sum_{j=0}^{d-1}\abs{\gamma_j}^2\cos\tilde{\lambda}_j$ with $\tilde{\lambda}_j:=\tfrac{\pi}{4}(1-\lambda_j)$. The normalized post-measurement system state for outcome $k \in \{0,1\}$ is [Eq.~(\ref{eq:output_state})]
\begin{eqnarray}
\ket{\varphi_{\mathrm{out},k}}=\frac{\bra{k}\hat{T}(\boldsymbol\theta)\ket{\Psi(\boldsymbol\theta)}}{\sqrt{p_k}}, \quad p_k=\frac{1}{2}.
\end{eqnarray}

Then, we present the following proposition and provide its proof:
\begin{proposition}[Ground-state amplitude amplification]
\label{prop:gsAA}
Assume $0 < \Gamma_{\ket{\alpha(\boldsymbol\theta)}} < 1$. For each $k\in\{0,1\}$, the step $\mathrm{({\bf A.1})}$ strictly increases the ground-state amplitude:
\begin{eqnarray}
\delta_{\lambda_0} &:=& \Gamma_{\ket{\varphi_{\mathrm{out},k}}} - \Gamma_{\ket{\alpha(\boldsymbol\theta)}} = 4 \abs{\mathcal{W}} \chi \xi_0 \abs{\gamma_0}^2 \nonumber \\
  &\ge& c_\star \Gamma_{\ket{\alpha(\boldsymbol\theta)}} \bigl(1-\Gamma_{\ket{\alpha(\boldsymbol\theta)}}\bigr) > 0,
\label{eq:gain_exact}
\end{eqnarray}
where 
\begin{eqnarray}
\begin{array}{l}
\chi := 4\abs{\mathcal{W}}^2 - 1, \\[2pt]
\xi_j := \abs{\mathcal{W}} - \cos\tilde{\lambda}_j, \\[2pt]
c_\star := 4\cos\tilde{\lambda}_0 \bigl( 4\cos^2\tilde{\lambda}_0 - 1 \bigr) \bigl(\cos\tilde{\lambda}_1 - \cos\tilde{\lambda}_0 \bigr) > 0.
\end{array}
\label{eq:cstar}
\end{eqnarray}
\end{proposition}

\begin{proof}---A direct calculation (expanding in the eigenbasis and using Eq.~(\ref{eq:U}) and Eq.~(\ref{eq:T})) gives the coefficient update
\begin{eqnarray}
\gamma_j \Rightarrow \gamma'_j =
\left\{
\begin{array}{ll}
\bigl(4\abs{\mathcal{W}}^2-2 \mathcal{W}^\ast e^{i \omega \lambda_j}-1 \bigr) \gamma_j,& k=0, \\[2pt]
\bigl(4\abs{\mathcal{W}}^2-2 i\mathcal{W}^\ast e^{-i \omega \lambda_j}-1 \bigr) \gamma_j,& k=1,
\end{array}
\right.
\label{eq:coef_update}
\end{eqnarray}
where $\abs{\gamma'_j}^2=\bigl( 1+4\abs{\mathcal{W}}\chi\xi_j \bigr) \abs{\gamma_j}^2$. Thus, we get $\delta_{\lambda_0} = 4\abs{\mathcal{W}}\chi\xi_0 \abs{\gamma_0}^2$. Here, the normalization follows from 
\begin{eqnarray}
\sum_j \xi_j \abs{\gamma_j}^2 = \abs{\mathcal{W}} - \sum_j \abs{\gamma_j}^2\cos\tilde{\lambda}_j=0.
\end{eqnarray}
Since $\lambda_0 < \lambda_1 \le \cdots \le \lambda_{d-1}$, we have 
\begin{eqnarray}
\cos\tilde{\lambda}_0 \le \abs{\mathcal{W}} \le \cos\tilde{\lambda}_{d-1},
\end{eqnarray}
and hence,
\begin{eqnarray}
\chi &\ge& 4\cos^2\tilde\lambda_0-1 \ge 1, \nonumber \\
\xi_0 &=& \sum_{j \ge 1} \abs{\gamma_j}^2 \bigl( \cos\tilde{\lambda}_j - \cos\tilde{\lambda}_0 \bigr) > 0.
\end{eqnarray}
Substituting the respective lower bounds for $\abs{\mathcal{W}}$ and $\chi$ establishes Eq.~(\ref{eq:gain_exact}); the inequality is valid for all nontrivial cases with $0 < \Gamma_{\ket{\alpha(\boldsymbol\theta)}} < 1$.
\end{proof}

We clarify that the result does not assume any particular expressivity of $\hat{A}(\boldsymbol\theta)$ beyond defining the current trial state; in fact, expressivity matters in ({\bf A.2}).

{\em Learn: state learning and stability.}---Given $\ket{\varphi_{\mathrm{out},k}}$ from ({\bf A.1}), the next trial is obtained by solving the state-learning problem:
\begin{eqnarray}
\boldsymbol\theta'=\underset{\boldsymbol\theta \in \Theta}{\mathrm{opt}} \mathcal F(\boldsymbol\theta),
\quad 
\ket{\alpha(\boldsymbol\theta')}=\hat{A}(\boldsymbol\theta')\ket{\mathbb 0} \simeq \ket{\varphi_{\mathrm{out},k}},
\label{eq:SL_prog}
\end{eqnarray}
with $\mathcal F$ an overlap/distance-based objective estimated from suitable observables; for example, maximize $\abs{\braket{\alpha(\boldsymbol\theta)}{\varphi_{\mathrm{out},k}}}^2$ or minimize the trace distance $D(\hat{\rho}_{\boldsymbol\theta}, \hat{\rho}_{\mathrm{out},k})$ where
\begin{eqnarray}
D=\frac{1}{2} \bigl\| \hat{\rho}_{\boldsymbol\theta} - \hat{\rho}_{\mathrm{out},k} \bigr\|_1.
\end{eqnarray}
Because the target is a \emph{fixed pure state} rather than an energy functional, ({\bf A.2}) avoids explicit exploration of rugged energy landscapes and admits well-studied smooth barrier or trust-region designs, e.g., using 
\begin{eqnarray}
\mathcal{F} =D^2 - \log(1-D^2)~\text{on}~D \in [0,1)
\label{eq:obj_fun}
\end{eqnarray}
for conditioning~\cite{Forst2010}. 

The key property needed for convergence is stability of the ground-state weight under learning error. Thus, we provide the following lemma:
\begin{lemma}[Stability under learning error]
\label{lem:stability}
Let $\hat{\rho}_{\mathrm{out},k}=\ket{\varphi_{\mathrm{out},k}}\bra{\varphi_{\mathrm{out},k}}$ and $\hat{\rho}_{\boldsymbol\theta'}=\ket{\alpha(\boldsymbol\theta')}\bra{\alpha(\boldsymbol\theta')}$. Then, for the projector $\hat{\Pi}_0 = \ket{\lambda_0}\bra{\lambda_0}$,
\begin{eqnarray}
\abs{\delta_{\lambda_0}} = \abs{\mathrm{Tr} \bigl[\hat{\Pi}_0 \bigl( \hat{\rho}_{\mathrm{out},k} - \hat{\rho}_{\boldsymbol\theta'} \bigr) \bigr]} \le D(\hat{\rho}_{\mathrm{out},k}, \hat{\rho}_{\boldsymbol\theta'}).
\label{eq:stability}
\end{eqnarray}
\end{lemma}

\begin{proof}---For any POVM element $0 \le \hat{\Pi} \le \hat{\openone}$, the following holds:
\begin{eqnarray}
\abs{\tr{\bigl[ \hat{\Pi} \bigl( \hat{\rho} - \hat{\rho}' \bigr) \bigr]}} \le \frac{1}{2} \bigl\| \hat{\rho} - \hat{\rho}' \bigr\|_1 = D(\hat{\rho}, \hat{\rho}').
\end{eqnarray}
Taking $\hat{\Pi}=\hat{\Pi}_0$ gives Eq.~(\ref{eq:stability}).
\end{proof}

\begin{figure}[t]
\includegraphics[width=0.46\textwidth]{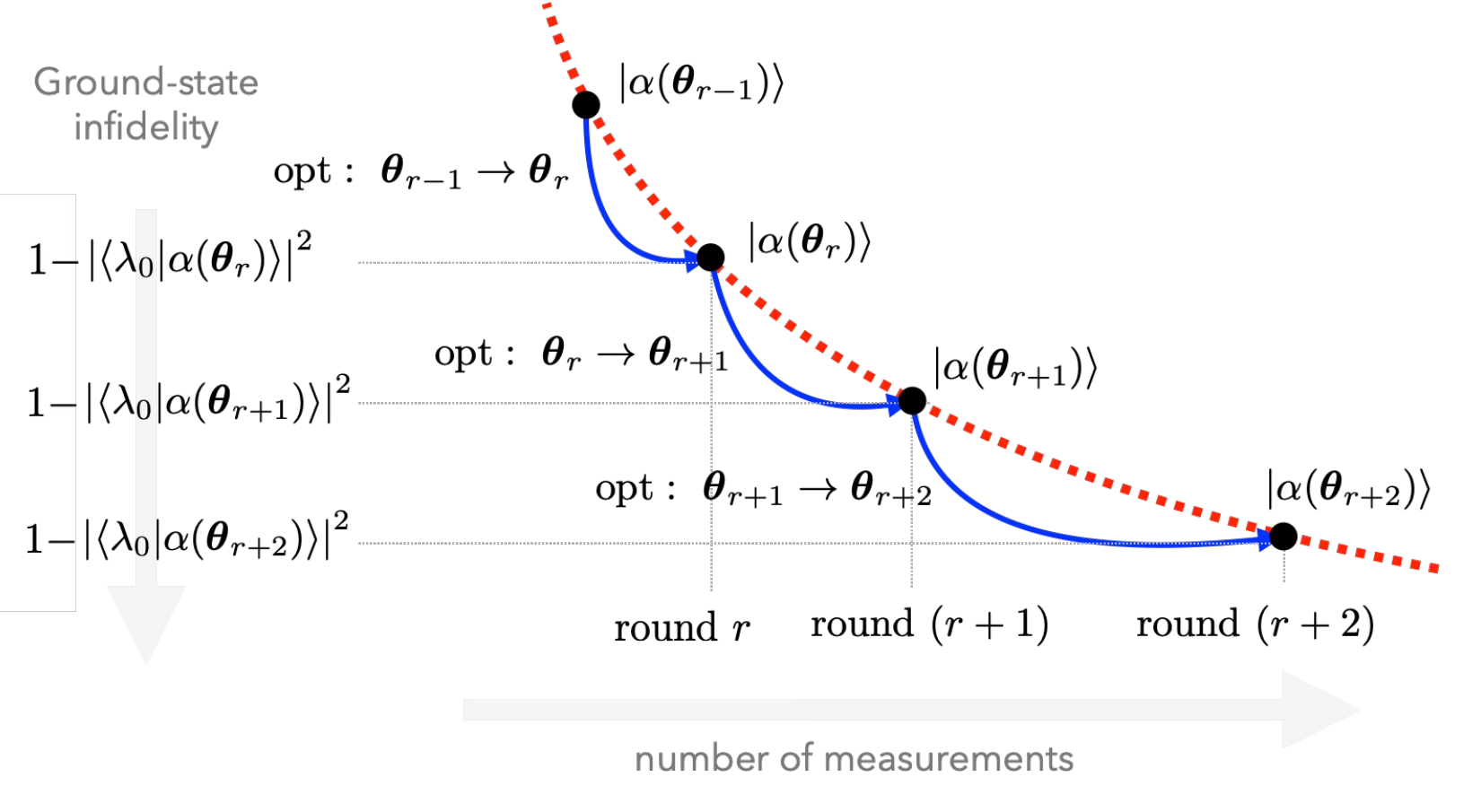}
\caption{The predicted behavior of converging to the ground state in our QAAE. Let us start by a system output state $\ket{\varphi_{\text{out},k}(r)}$ at any $r$-th round, which is closer to the ground state than those from the previous rounds. At this round, the state learning is completed and the control parameter vector is updated as $\boldsymbol\theta_{r} \to \boldsymbol\theta_{r+1}$, and the PQC states $\ket{\alpha(\boldsymbol\theta_{r+1})}$ is prepared (see the blue solid line). By iterating these processes, the PQC states evolves into the ground state, which is depicted as a trajectory along the red dashed line.}
\label{fig:conv_behavior}
\end{figure}

Combining {\bf Proposition~\ref{prop:gsAA}} with {\bf Lemma~\ref{lem:stability}} yields a per-round bound. Let $g_r:=\Gamma_{\ket{\alpha(\boldsymbol\theta_r)}}$ and $\varepsilon_r:=D\left(\ket{\alpha(\boldsymbol\theta_{r+1})}, \ket{\varphi_{\mathrm{out},k}(r)}\right)$ denote the learning error at round $r$. Then,
\begin{eqnarray}
g_{r+1}\ \ge\ g_r \ +\ c_\star\,g_r(1-g_r)\ -\ \varepsilon_r.
\label{eq:meta}
\end{eqnarray}
In particular: (i) Exact learning (i.e., $\varepsilon_r=0$) gives a strictly increasing, bounded sequence with logistic-type gain, hence $g_r$ approaches $1$. (ii) Approximate learning preserves monotonicity whenever $\varepsilon_r<c_\star\,g_r(1-g_r)$; if $\varepsilon_r\to0$ (or $\sum_r \varepsilon_r < \infty$), the sequence still converges to $1$. Figure~\ref{fig:conv_behavior} illustrates the predicted trajectory. 

{\em Convergence via amplify--learn.}---The amplify step provides a state-independent improvement mechanism ({\bf Proposition~\ref{prop:gsAA}}) whose strength is set by spectral constants of $\hat{H}$ via $c_\star > 0$, while the learn step translates this improvement into the next round within the chosen ansatz class, with the robustness quantified by Eq.~(\ref{eq:stability}). Note that our QAAE operates outside the variational paradigm: it neither navigates rugged Hamiltonian energy landscapes nor depends on energy gradients. Robustness to imperfect learning is quantified by the stability bound in Eq.~(\ref{eq:stability}), which, combined with the per‑round overlap gain, yields the recursion Eq.~(\ref{eq:meta}) (with $g_r=\Gamma_{\ket{\alpha(\boldsymbol\theta_r)}}$ and $\varepsilon_r=D(\hat{\rho}_{\boldsymbol\theta_{r+1}}, \hat{\rho}_{\text{out},k})$). Under the standard assumptions and faithful (or asymptotically faithful) learning, this establishes a \emph{provably monotone and convergent} route to the ground state.

\subsection{Circuit Depth}  

We bound the circuit depth per QAAE round. Each round comprises: (i) trial‑state preparation $\hat{A}(\boldsymbol\theta)$, (ii) the Householder reflection $\hat{R}(\boldsymbol\theta)$ about $\ket{\Psi(\boldsymbol\theta)}$, and (iii) the controlled short‑time evolution $\hat{U}$. Denote their depths by $\mathcal{D}_A$, $\mathcal{D}_R$, and $\mathcal{D}_U$, respectively.

\begin{proposition}[Per‑round circuit depth]
\label{prop:depth}
For a normalized Hamiltonian $\hat{H}$ and $\hat U$ in Eq.~(\ref{eq:U}) with fixed $\omega$, one QAAE round has circuit depth
\begin{eqnarray}
\mathcal{D}_{\mathrm{round}} \le \mathcal{D}_A + 2\bigl(\mathcal{D}_R+\mathcal{D}_U \bigr),
\label{eq:rounddepth}
\end{eqnarray}
with
\begin{equation}
\mathcal{D}_R \le 2\mathcal{D}_A + \mathcal{O}(q),  \quad  \mathcal{D}_U = \mathcal{O} \bigl( \mathrm{poly}(q) \varepsilon_H^{-\frac{1}{2\kappa}} \bigr),
\label{eq:DRDU}
\end{equation}
where $q$ is the number of system qubits, $\varepsilon_H$ is the target simulation accuracy for $e^{-i\omega\hat{H}}$, and $2\kappa$ is the order of a Trotter-Suzuki product formula (other short‑time simulation primitives may be used). Consequently,
\begin{equation}\label{eq:rounddepth_simple}
\mathcal{D}_{\mathrm{round}} \le 5\mathcal{D}_A\ + 2\mathcal{D}_U + \mathcal{O}(q).
\end{equation}
\end{proposition}

\begin{proof}[Proof sketch]---Using the canonical construction $\hat{R}(\boldsymbol\theta)= \bigl(\mathrm{Hadamard}\otimes\hat A(\boldsymbol\theta)\bigr) \hat{R}_0 \bigl(\mathrm{Hadamard}\otimes\hat A(\boldsymbol\theta)\bigr)^\dagger$ with $\hat{R}_0 = \hat{\openone} - 2\ket{\mathbb{0}}\bra{\mathbb{0}}$ (as illustrated in Fig.~\ref{fig:schematic}(b)), the reflection depth is that of $\hat{A}(\boldsymbol\theta)$, its adjoint $\hat{A}(\boldsymbol\theta)^\dagger$, an ancilla Hadamard, and a $(q{+}1)$‑controlled phase (a relative‑phase multi‑controlled $Z$). Here, standard decompositions give depth $\mathcal{O}(q)$ for this multi‑controlled gate; with an extra ancilla, $\mathcal{O}(\log q)$ is possible while keeping width $\mathcal{O}(q)$~\cite{Saeedi2013}. Hence, $\mathcal{D}_R\le 2\mathcal{D}_A+\mathcal{O}(q)$.

For $e^{-i\omega\hat{H}}$ with fixed $\omega=\tfrac{\pi}{4}$), a $2\kappa$‑th order product formula over $\hat{H}=\sum_{l=1}^L \hat h_l$ yields the step cost $\mathcal{O}(L)$ and step count $\nu=\mathcal{O}\bigl((L \omega)^{1+{1}/{2\kappa}} \varepsilon_H^{-{1}/{2\kappa}} \bigr)$~\cite{Berry2007}. For typical local models $L=\mathrm{poly}(q)$, so $\mathcal{D}_U = \mathcal{O}\bigl(\mathrm{poly}(q) \varepsilon_H^{-{1}/{2\kappa}}\bigr)$. The controlization adds only constant‑factor overhead for Pauli‑string terms and can be scheduled with pre/post controlled Paulis~\cite{Dong2022}. 

One round of QAAE executes state preparation once and the pattern $\hat{R}\hat{U}\hat{R}\hat{U}^\dagger$, giving Eq.~(\ref{eq:rounddepth}). Substituting Eq.~(\ref{eq:DRDU}) yields Eq.~(\ref{eq:rounddepth_simple}).
\end{proof}

{\em Hardware‑conscious design.}---Now we remark the following. (i) Ansatz modularity: The bound in Eq.~(\ref{eq:rounddepth_simple}) separates the ansatz cost $\mathcal{D}_A$ from Hamiltonian simulation $\mathcal{D}_U$. Chemistry‑inspired circuits (e.g., UCC‑type) and hardware‑efficient layouts both fit this template; thus QAAE inherits their depth‑scaling properties~\cite{Ostaszewski2021}. (ii) Parallel scheduling: For local Hamiltonians, the Pauli terms partition into commuting layers; with standard graph colorings of the interaction hyper-graph, one obtains constant (model‑dependent) parallel depth per Trotter step, improving the constant in $\mathrm{poly}(q)$ without changing the asymptotics. Controlled‑Pauli ``wraps'' can be hoisted and canceled across layers to reduce two‑qubit gate count~\cite{Dong2022}. (iii) Beyond Trotter: LCU/QSP‑based short‑time simulation achieves asymptotically better accuracy dependence but typically requires block‑encoding and additional ancillas; for NISQ/early‑FTQC contexts, product formulas often yield lower overheads~\cite{Babbush2021,Campbell2021,Zhang2022,Layden2022,Sun2024}. Either choice plugs into Eq.~(\ref{eq:rounddepth}) by substituting the corresponding $\mathcal{D}_U$. (iv) Controlled‑$Z$ fan‑in: The $\mathcal{O}(q)$ term in Eq.~(\ref{eq:DRDU}) stems from the $(q+1)$‑controlled phase in $\hat{R}_0$; the depth can be reduced to $\mathcal{O}(\log q)$ with an ancilla ladder at the cost of a small qubit overhead~\cite{Saeedi2013}. (v) Error budgeting: Because QAAE's per‑round gain is set by spectral constants ({\bf Proposition~\ref{prop:gsAA}}), the simulation accuracy $\varepsilon_H$ needs only be tight enough that Trotter error remains below this gain; thus $\mathcal{D}_U$ can be chosen by a simple budget rule rather than by the stringent precision targets common in phase‑estimation pipelines. 

Overall, Eqs.~(\ref{eq:rounddepth})--(\ref{eq:rounddepth_simple}) show that the amplify--learn loop admits a \emph{polynomial} per‑round circuit depth governed by the chosen ansatz and short‑time simulation primitive, underscoring QAAE's hardware compatibility.

\section{Applications}\label{Sec:4}

\subsection{Motivating example: a two-level system Hamiltonian} 

To isolate and validate the core mechanism of QAAE, we consider the physically universal qubit Hamiltonian
\begin{eqnarray}
\hat H_{\mathrm{2L}}=\alpha\hat{\openone} + \mathbf{r} \cdot \hat{\boldsymbol\sigma} \quad\text{with}\quad \hat{\boldsymbol\sigma}=(\hat\sigma_x,\hat\sigma_y,\hat\sigma_z)^T,
\end{eqnarray}
whose spectrum is $\lambda_{0,1}= \alpha \mp \abs{\mathbf{r}}$ with the energy gap $d = \lambda_1-\lambda_0=2\abs{\mathbf{r}}$. Equivalently,
\begin{equation}
\hat H_{\mathrm{2L}}= \lambda_0 \hat{\openone} + d \ket{\lambda_1}\bra{\lambda_1},
\end{equation}
where $\{\ket{\lambda_0}, \ket{\lambda_1}\}$ are unknown. This archetypal two‑level model cleanly probes whether a single QAAE round coherently amplifies the ground‑state weight as predicted by {\bf Proposition~\ref{prop:gsAA}}, without confounding effects from complex many‑body structure. Crucially, we deliberately disable the state-learning here: the next trial is obtained by characterizing the post‑measurement output via a small set of observables, so that any improvement can be attributed purely to the amplitude amplification.

For a general $q$‑qubit ansatz, we expand
\begin{eqnarray}
\hat{\rho}(\boldsymbol\theta)=\ket{\alpha(\boldsymbol\theta)}\bra{\alpha(\boldsymbol\theta)} = \frac{1}{2^q}\sum_{\mathbf{v} \in \mathcal{I}} C_{\mathbf{v}}(\boldsymbol\theta)\bigotimes_{j=1}^q \hat{\sigma}_{v_j},
\label{eq:PauliExp}
\end{eqnarray}
where $v_j \in \{0,1,2,3\}$, with $\hat{\sigma}_0 = \hat{\openone}$, $\hat{\sigma}_1 = \hat{\sigma}_x$, $\hat{\sigma}_2 = \hat{\sigma}_y$, and $\hat{\sigma}_3 = \hat{\sigma}_z$. The index string $\mathbf{v} = v_1 v_2 \cdots v_q$ specifies a $q$-qubit Pauli string, and the index set $\mathcal{I} \subseteq \{ 0,1,2,3 \}^q$ defines the subset of Pauli strings contributing to the expansion: i.e., $\{\bigotimes_j \hat{\sigma}_{v_j}\}_{\mathbf{v} \in \mathcal{I}}$, where $|\mathcal I|\ll 4^q$ in practice. The coefficients $C_{\mathbf{v}}(\boldsymbol{\theta})$ depend on the parameters $\boldsymbol{\theta}$ and encode the structure of the ansatz-generated state in the Pauli basis. In the present single‑qubit case ($q{=}1$), we restrict to the $x$–$z$ meridian and use
\begin{eqnarray}
\hat{\rho}(\theta)=\frac{1}{2}\bigl( \hat{\openone} + C_x(\theta)\hat{\sigma}_x + C_z(\theta)\hat{\sigma}_z \bigr),
\end{eqnarray}
with $C_x(\theta)=\sin\theta$ and $C_z(\theta)=\cos\theta$. Each round measures $\{\hat{\sigma}_x, \hat{\sigma}_z\}$ on the output state $\ket{\varphi_{\mathrm{out},k}}$ of (\textbf{A.1}), forming $\mathbf{c}=\bigl(c_x=\expt{\hat{\sigma}_x}, c_z=\expt{\hat{\sigma}_z}\bigr)$. We then project to the Bloch sphere by normalization $\tilde{\mathbf{c}}={\mathbf{c}}/{\abs{\mathbf{c}}}$ (a maximum‑likelihood pure‑state update under isotropic noise) and set the next trial by $\theta'=\mathrm{atan2}(\tilde{c}_x, \tilde{c}_z)$, thereby re‑encoding the amplified state without evaluating the variational gradients.

\begin{figure}[t]
\includegraphics[width=0.40\textwidth]{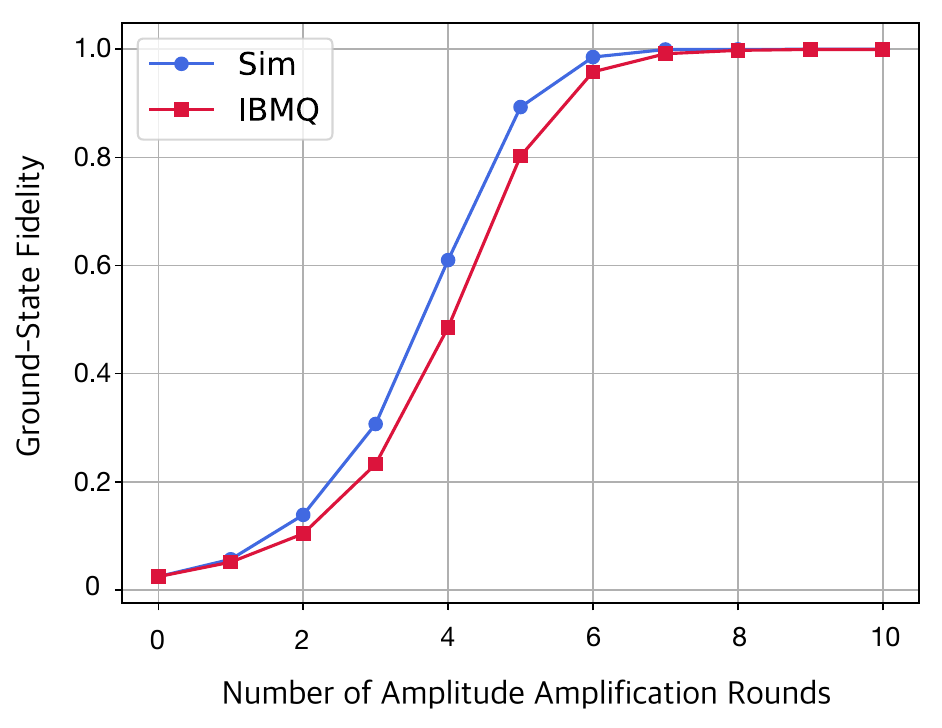}
\caption{Ground‑state fidelity vs. number of QAAE rounds for the two‑level Hamiltonian. Both Qiskit emulation and \texttt{ibm\_yonsei} experiments show monotone improvement consistent with {\bf Proposition~\ref{prop:gsAA}}.}
\label{fig:f_ex1}
\end{figure}

\begin{figure}[t]
\includegraphics[width=0.46\textwidth]{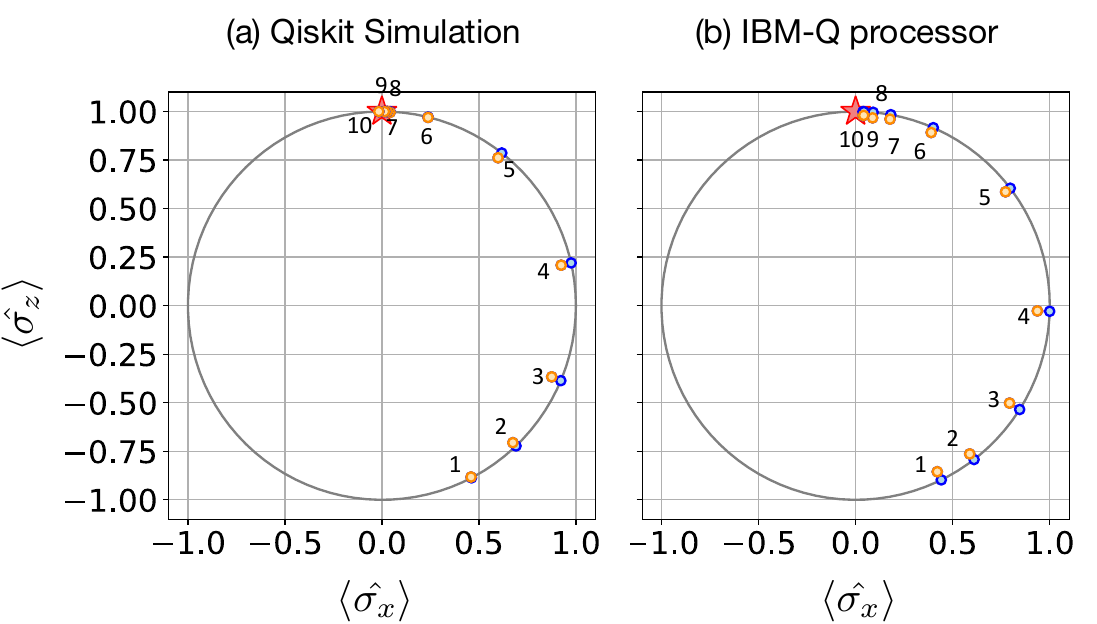}
\caption{Expectation values $\expt{\hat{\sigma_x}}$ and $\expt{\hat{\sigma_z}}$ from (a) Qiskit emulation and (b) \texttt{ibm\_yonsei} experiments (orange), and corresponding normalized Bloch vectors used for re‑encoding (blue). The red star marks the ground state. The normalization projects noisy Bloch vectors back to purity, mitigating isotropic‑type noise and improving update robustness.}
\label{fig:comparison_IBM}
\end{figure}

\begin{figure}[t]
\center
 \includegraphics[width=0.46\textwidth]{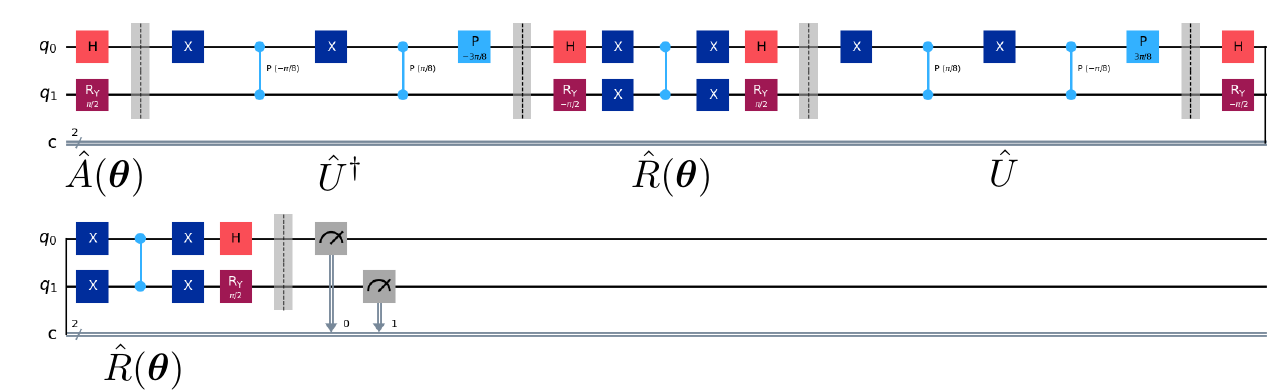}
 \caption{Single‑ancilla QAAE circuit for the two‑level Hamiltonian. One round consists of the reflection--evolution--reflection--inverse‑evolution pattern of Sec.~\ref{Sec:2}, followed by ancilla measurement and re‑encoding of the amplified pure state.}
\label{fig:f_ex1_circuit}
\end{figure}

We then implement the single‑ancilla QAAE round on the IBMQ cloud processor ``\texttt{ibm\_yonsei}'' and compare to a Qiskit‑based local emulator. The initial trial has very low ground‑state fidelity (lower than $0.024$ on average). Figure~\ref{fig:f_ex1} shows that the fidelity increases steadily toward unity as the number of amplification rounds grows, confirming {\bf Proposition~\ref{prop:gsAA}} on hardware. In Fig.~\ref{fig:comparison_IBM}, orange dots are raw expectation values $(\expt{\hat{\sigma}_x}, \expt{\hat{\sigma}_z})$ from (a) emulation and (b) experiment; blue dots are the normalized Bloch vectors used for re‑encoding. Their close alignment indicates that the simple normalization step acts as an effective denoiser against approximately isotropic noise, stabilizing the round‑to‑round parameter update. The ground state (red star) is approached monotonically under repeated amplification. Circuit‑level details for this one‑qubit instance are shown in Fig.~\ref{fig:f_ex1_circuit}. 

Although $\hat H_{\mathrm{2L}}$ acts on a single qubit, it is the canonical normal form of any effective two‑level subspace (e.g., avoided crossings, spin‑$1/2$ in a field). Here, the energy gap $d$ sets the constant governing the per‑round improvement in {\bf Proposition~\ref{prop:gsAA}}, and the experiment demonstrates that QAAE's amplitude amplification translates into observable, monotone fidelity gains on real hardware---without invoking the variational principle or exploring an energy landscape. The same observable‑driven re‑encoding idea scales: for larger $q$, one may replace $\{\hat{\sigma}_x, \hat{\sigma}_z\}$ by a Pauli set as in Eq.~(\ref{eq:PauliExp}) and exploit entangled measurements~\cite{Hamamura2020} or classical‑shadow estimation~\cite{Huang2020} to keep the circuit cost practical.

\subsection{Hydrogen molecule} 

\begin{figure*}[t]
	\includegraphics[width=1.00\textwidth]{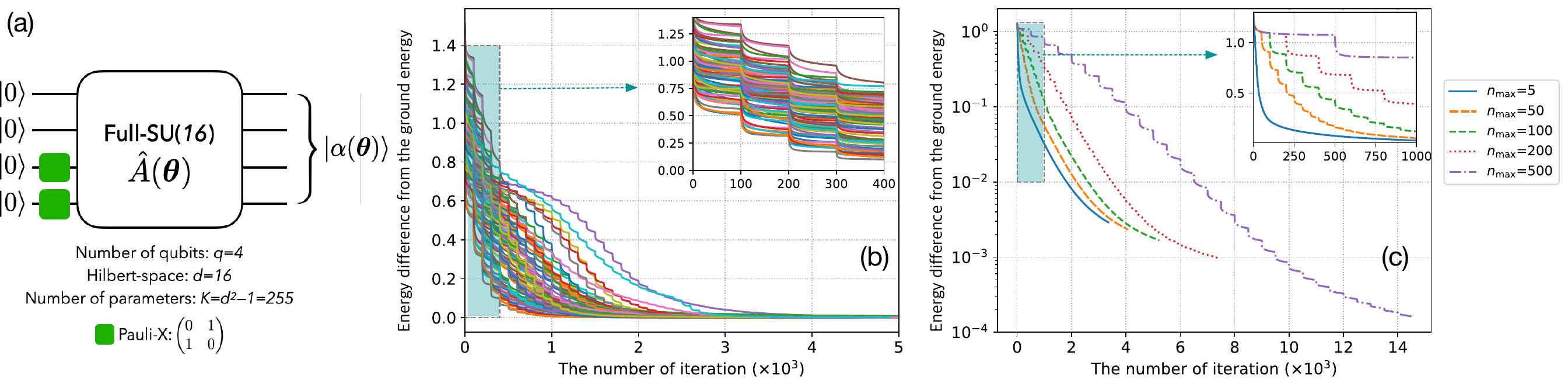}
	\caption{QAAE for $\mathrm{H}_2$ with a full SU($16$) ansatz. (a) The ansatz circuit $\hat{A}(\boldsymbol\theta)$ is a full SU($16$) unitary, enabling exploration of the entire $16$-dimensional Hilbert space ($K=16^2-1=255$ parameters). (b) QAAE over $100$ random initializations (applied to the $4$‑qubit Hartree-Fock reference $\ket{0011}$) exhibits robust, monotone improvement consistent with the analysis in Sec.~\ref{Sec:3}. (c) Effect of limiting state‑learning effort per round to $n_{\max} \in \{ 500, 200, 100, 50, 5\}$: smaller $n_{\max}$ accelerates early‑stage progress but requires more rounds to reach high‑precision estimates.}
	\label{fig:h2_su16}
\end{figure*}

\begin{figure*}[t]
	\includegraphics[width=1.00\textwidth]{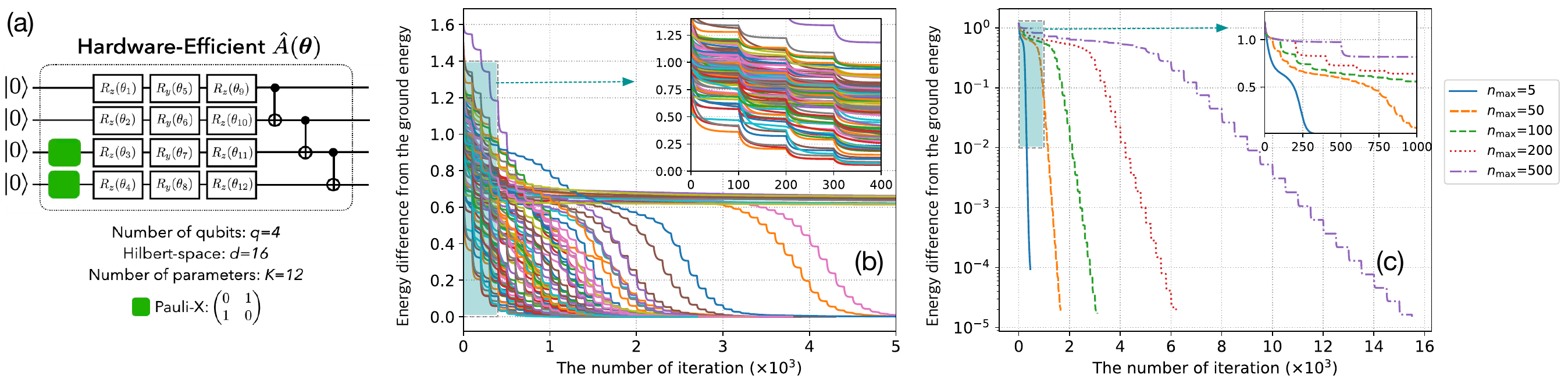}
	\caption{QAAE for $\mathrm{H}_2$ with a hardware‑efficient ansatz. (a) Two‑layer hardware‑efficient circuit $\hat{A}(\boldsymbol\theta)$ (single‑qubit rotations + CNOT entanglers); $K=24$ parameters. (b) Over $100$ random seeds with $n_{\max}=100$, about $90\%$ reach the ground state; the remaining $10\%$ stall due to limited expressivity and state-learning. (c) Varying $n_{\max} \in \{ 500, 200, 100, 50, 5 \}$ shows fast early‑round gains for small $n_{\max}$, with the final accuracy less sensitive to $n_{\max}$ than in the SU($16$) case.}
	\label{fig:h2_HE}
\end{figure*}

As a first molecular application, we target the ground state of $\mathrm{H}_2$. Here, we employ a four‑qubit Hamiltonian $\hat{H}_{\mathrm{H}_2}$ obtained from second‑quantized electronic structure in STO‑3G followed by a Jordan-Wigner mapping~\cite{McArdle2020}; the internuclear separation is fixed to $0.7 \text{\AA}$. Within QAAE, ({\bf A.1}) implements the amplitude‑amplification round using $\hat{A}(\boldsymbol\theta)$, $\hat{U}$, $\hat{R}_0$, and $\hat{T}(\boldsymbol\theta)$; ({\bf A.2}) performs the state-learning to re‑encode the amplified target state. 

{\em Learning objective and halting.}---For ({\bf A.2}), we adopt stochastic gradient descent on the distance‑based objective $\mathcal{F}(D)$, given in Eq.~(\ref{eq:obj_fun}). Here, $D=\tfrac{1}{2} \bigl\| \hat{\rho}_{\mathrm{out},k} - \hat{\rho}_{\boldsymbol\theta'} \bigr\|_1$, which is the trace distance between the amplified target $\hat{\rho}_{\mathrm{out},k}=\ket{\varphi_{\mathrm{out},k}}\bra{\varphi_{\mathrm{out},k}}$ and the next‑round trial $\hat{\rho}_{\boldsymbol\theta'}=\ket{\alpha(\boldsymbol\theta')}\bra{\alpha(\boldsymbol\theta')}$. The barrier term in Eq.~(\ref{eq:obj_fun}) regularizes large‑distance regions while keeping a well‑conditioned minimum at the target~\cite{Forst2010}. The procedure halts at round $r$ when
\begin{eqnarray}
\Delta_{r-1,r} = \abs{E_{r} - E_{r-1}} \le \epsilon_H,
\end{eqnarray}
where $E_r = \bra{\alpha(\boldsymbol\theta_r)}\hat{H}_{\mathrm{H}_2}\ket{\alpha(\boldsymbol\theta_r)}$. Here, $\varepsilon_H=10^{-6}$. Note that the energy $E_r$ here is used only as a stopping certificate; it is not the objective optimized within the loop.

{\em Full SU($d$) ansatz (expressivity upper bound).}---We adopt the ansatz operation $\hat{A}(\boldsymbol\theta) \in \mathrm{SU}(d)$ with $d=16$. Although such circuits are impractical for large‑scale chemistry, they provide a clean lens on ({\bf A.1})–({\bf A.2}) dynamics without expressivity bottlenecks. Initializing from the Hartree-Fock determinant, numerical runs across $100$ random seeds show that the ground‑state weight in $\ket{\alpha(\boldsymbol\theta)}$ increases round‑by‑round, in line with {\bf Proposition~\ref{prop:gsAA}} and the logistic‑type bound derived in Sec.~\ref{Sec:3}; see Fig.~\ref{fig:h2_su16}(b). We then cap the learning budget per round at $n_{\max}$ gradient steps. As shown in Fig.~\ref{fig:h2_su16}(c), small $n_{\max}$ yields a greedy regime---rapid initial energy drop due to repeated re‑encoding of partially learned states---followed by a plateau that dissolves once additional learning steps are allowed. This quantifies a practical speed-vs-accuracy trade‑off: early rounds can prioritize speed (small $n_{\max}$), while later rounds allocate more learning to sharpen the estimate.

{\em Hardware‑efficient ansatz (NISQ‑oriented).}---We next adopt a two‑layer hardware‑efficient circuit (single‑qubit rotations and CNOT entanglers; control parameters $K=24$), which is natural for NISQ devices but cannot span the full Hilbert space. In this setting, ({\bf A.1}) still supplies a state‑independent overlap gain, while ({\bf A.2}) may fail to perfectly re‑encode the amplified target when the target leaves the ansatz manifold. Numerically, with $n_{\max}=100$, approximately $90\%$ of $100$ random seeds converge to the ground state; the remaining $10\%$ display stagnation attributable to expressivity‑limited learning (see Fig.~\ref{fig:h2_HE}(b)). As in the SU($16$) study, reducing $n_{\max}$ accelerates early improvements (Fig.~\ref{fig:h2_HE}(c)), but here the terminal accuracy is comparatively less sensitive to $n_{\max}$, reflecting that the dominant limitation is expressivity rather than learning budget.

Across both ansatz families, QAAE maintains its non‑variational identity: the loop does not minimize $\bra{\Psi}\hat H\ket{\Psi}$ nor its gradients; it learns a fixed pure state produced by the coherent amplification. This avoids explicit exploration of rugged energy landscapes, while preserving hardware compatibility through short‑time simulation in $\hat{U}$ and modular control of $\mathcal{D}_A$. In practice, one can schedule $n_{\max}$ to emphasize fast progress early and high precision late, and select ansatz families (chemistry‑inspired vs. hardware‑efficient) according to the target system and device constraints.

\subsection{LiH molecule} 

As a second molecular test, we compute the potential energy curve (PEC) of lithium hydride (LiH)~\cite{Aspuru2005} and assess QAAE across bond lengths. Here, we use a $12$‑qubit second‑quantized Hamiltonian $\hat{H}_{\mathrm{LiH}}$ obtained in the STO‑3G basis with a Jordan–Wigner mapping (normalized as $\tfrac{1}{16}\hat{H}_{\mathrm{LiH}}$)~\cite{McArdle2020}. As in the previous $\mathrm{H}_2$ example, ({\bf A.1}) performs amplitude amplification and ({\bf A.2}) state-learning re‑encodes the amplified target in a chosen ansatz---here, unitary coupled‑cluster singles and doubles (UCCSD). The learning uses stochastic gradient descent on $\mathcal F(D)$ in Eq.~(\ref{eq:obj_fun}); to emulate tight hardware budgets we cap the learning effort per round at $n_{\max}=5$.

{\em Setup and protocol.}---For each bond length $R \in [1.0, 3.3] \text{\AA}$, we initialize the UCCSD circuit $\hat{A}(\boldsymbol\theta)$ on the Hartree-Fock reference. Each QAAE round executes one amplify step to generate $\ket{\varphi_{\mathrm{out},k}}$, measures the observables needed for $\mathcal{F}(D)$, and performs at most $n_{\max}=5$ learning steps to re‑encode the amplified target as $\ket{\alpha(\boldsymbol\theta')}$. The energy $E(R)=\bra{\alpha(\boldsymbol\theta)}\hat{H}_{\mathrm{LiH}}\ket{\alpha(\boldsymbol\theta)}$ is recorded for monitoring and halting, yet not the optimization objective within the loop.

\begin{figure}[t]
\centering
\includegraphics[width=0.46\textwidth]{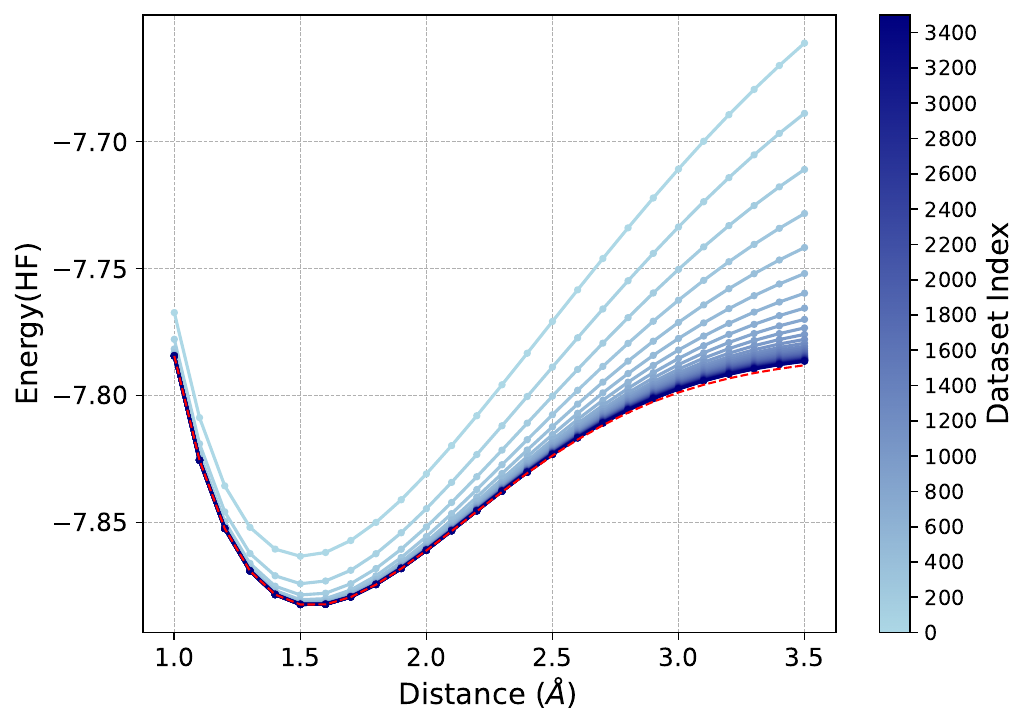}
\caption{LiH potential energy curve with UCCSD ansatz ($12$ qubits). Bond scan from $1.0 \text{\AA}$ to $3.3 \text{\AA}$. Blue traces show per‑round QAAE estimates (earliest rounds in lighter shades); the red dashed line is the exact ground‑state energy for our $12$‑qubit model. QAAE drives the PEC toward the exact curve under a small learning budget $n_{\max}=5$, illustrating landscape‑free convergence with a chemistry‑inspired ansatz.}
\label{fig:LiH}
\end{figure}

{\em Results and accuracy.}---Figure~\ref{fig:LiH} shows that the PEC (blue) converges monotonically toward the exact curve (red dashed) across the scan. The equilibrium bond length is recovered at $R=1.5 \text{\AA}$. With a total of $3{,}500$ predicted trial states across the scan, the energy error at equilibrium is $1.8 \times 10^{-5}$~Ha, while at $R=3.2 \text{\AA}$ (near dissociation) it is $1.6 \times 10^{-3}$~Ha. The dissociation energy extracted from the QAAE PEC is $0.0958$~Ha; for our $12$‑qubit model the exact value is $0.0942$~Ha, and the experimental reference is $0.0922$~Ha. These data indicate that even under a very small learning budget per round, QAAE's coherent amplification yields a quantitatively reliable PEC, approaching chemical‑accuracy scales near the dissociation tail.

UCCSD supplies chemically structured expressivity, while QAAE's amplify--learn loop provides directionality without energy‑landscape exploration. The combination achieves robust, hardware‑conscious convergence: short‑time simulation in $\hat{U}$ controls depth (Sec.~\ref{Sec:3}), and the limited learning budget trades a modest increase in rounds for stable, monotone PEC refinement.

\subsection{Ising model in transverse and longitudinal fields} 

Finally, for a more practical application, we cast the 1D longitudinal–transverse‑field Ising model (LTFIM),
\begin{eqnarray}
\hat{H}_\text{LTFIM} =  \sum_{j=1}^{N-1} \hat{\sigma}_z^{(j)} \hat{\sigma}_z^{(j+1)}+ \sum_{j=1}^{N} \bigl( g\hat{\sigma}_x^{(j)} + h \hat{\sigma}_z^{(j)} \bigr),
\end{eqnarray}
where $g$ and $h$ are transverse and longitudinal fields, respectively. The model is integrable at $h=0$ (via Jordan-Wigner) but becomes nonintegrable for $h \neq 0$~\cite{Lieb1961,Vsamaj2013}; its critical line connects $(g,h)=(1,0)$ to $(0,2)$~\cite{Ovchinnikov2003}. As such, LTFIM furnishes a stringent test bed that spans from exactly solvable to quantum‑chaotic regimes.

\begin{figure}[t]
\centering
\includegraphics[width=0.46\textwidth]{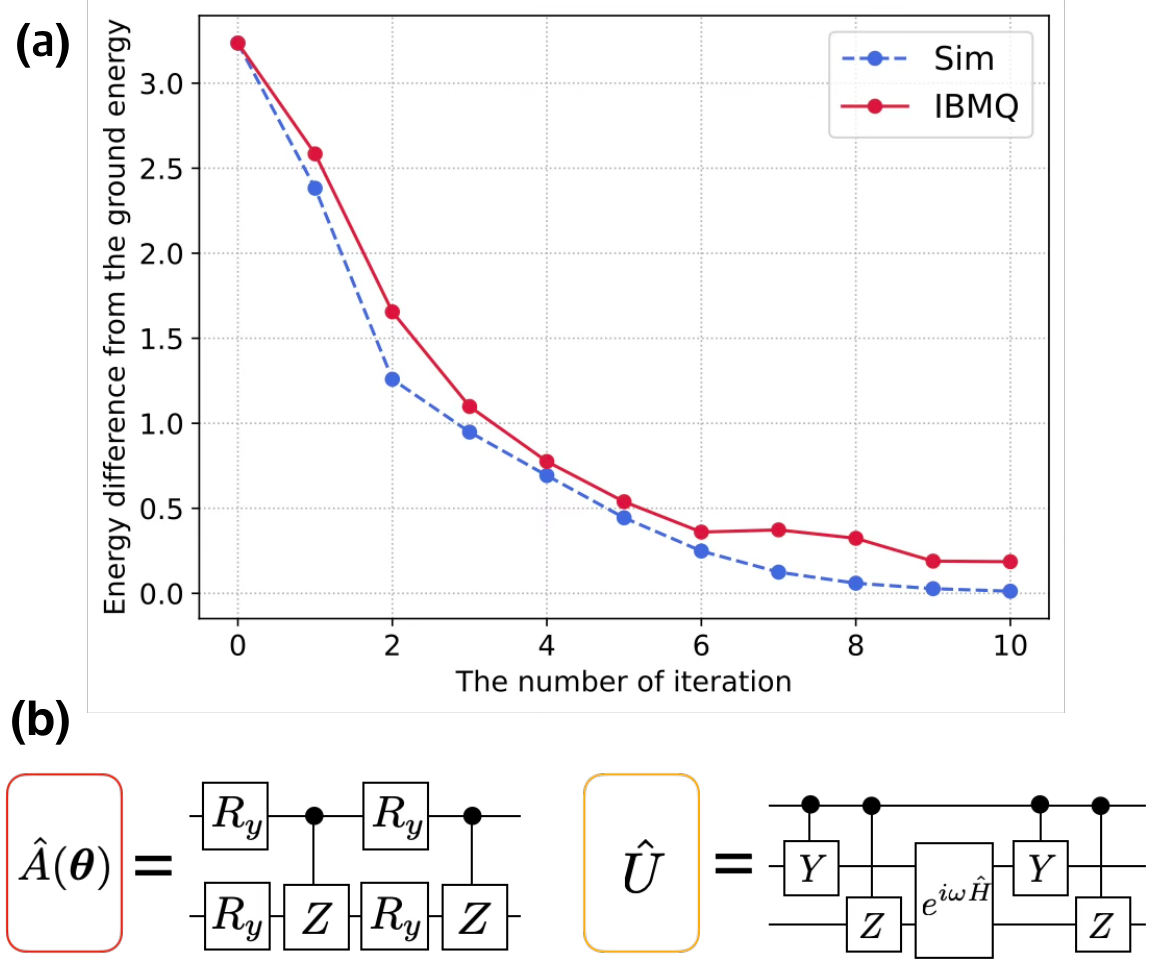}
\caption{Two‑qubit model on IBMQ hardware ($N=2$, and $(g, h)=(1,0)$). (a) Energy error (relative to the exact ground energy) vs. QAAE rounds with $n_{\max}=1$ learn‑updates per round: experiment (``\texttt{ibm\_yonsei}'' IBMQ processor) vs. numerical emulation agree and decrease monotonically. (b) The hardware efficient ansatz $\hat{A}(\boldsymbol\theta)$ and controlled evolution $\hat{U}$ circuit used in the implementation.}
\label{fig:TFIM}
\end{figure}

\begin{figure}[t]
\center
 \includegraphics[width=0.46\textwidth]{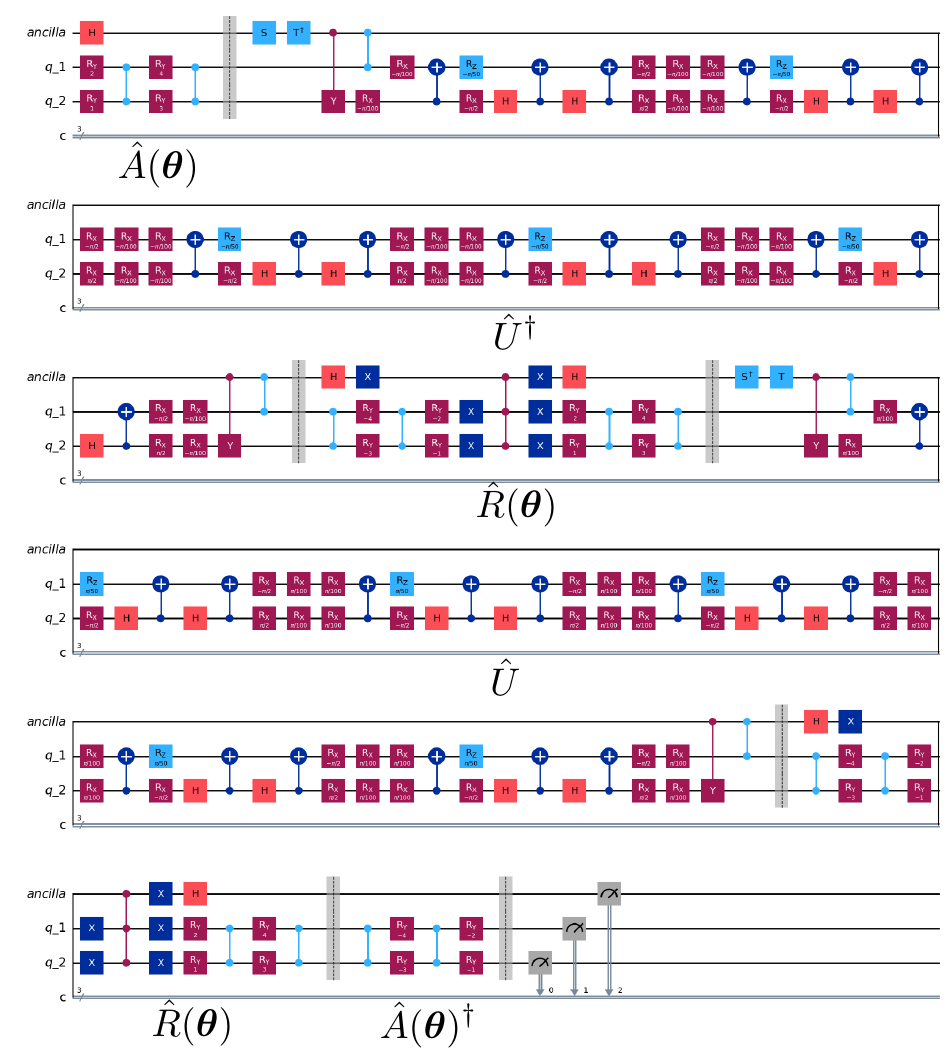}
 \caption{The per-round QAAE circuit for IBMQ processor experiment: $N=2$ and $(g, h)=(1,0)$.}
\label{fig:TFIM_circuit}
\end{figure}

{\em Two‑qubit hardware demonstration.}---We first perform a IBMQ cloud experiment on \emph{hardware} with $N=2$ ($g=1$ and $h=0$) using a single ancilla qubit to implement the QAAE rounds. The controlled short‑time evolution $\hat{U}$ is compiled by sandwiching a Trotterized $e^{-i\omega\hat H}$ between controlled Pauli operations $\bigl( \hat{\sigma}^{(1)}_y \otimes \hat{\sigma}^{(2)}_z \bigr)$, following the depth‑reduction scheme of Ref.~\cite{Dong2022} (see the circuit in Fig.~\ref{fig:TFIM}(b) and Fig.~\ref{fig:TFIM_circuit}). Here, five second‑order Trotter steps are used together with standard Qiskit optimization tool. Starting from a hardware‑efficient ansatz $\hat{A}(\boldsymbol\theta)$ with an initial $\boldsymbol\theta$ and a small initial ground‑state overlap, the QAAE energy error decreases monotonically with the number of rounds, in close agreement with noiseless emulations (see Fig.~\ref{fig:TFIM}(a)). This validates, \emph{on real device} (``\texttt{ibm\_yonsei}'' IBMQ processor), that the amplify--leaning loop provides the coherent improvement. 

\begin{figure*}[t]
\centering 
\includegraphics[width=1.00\textwidth]{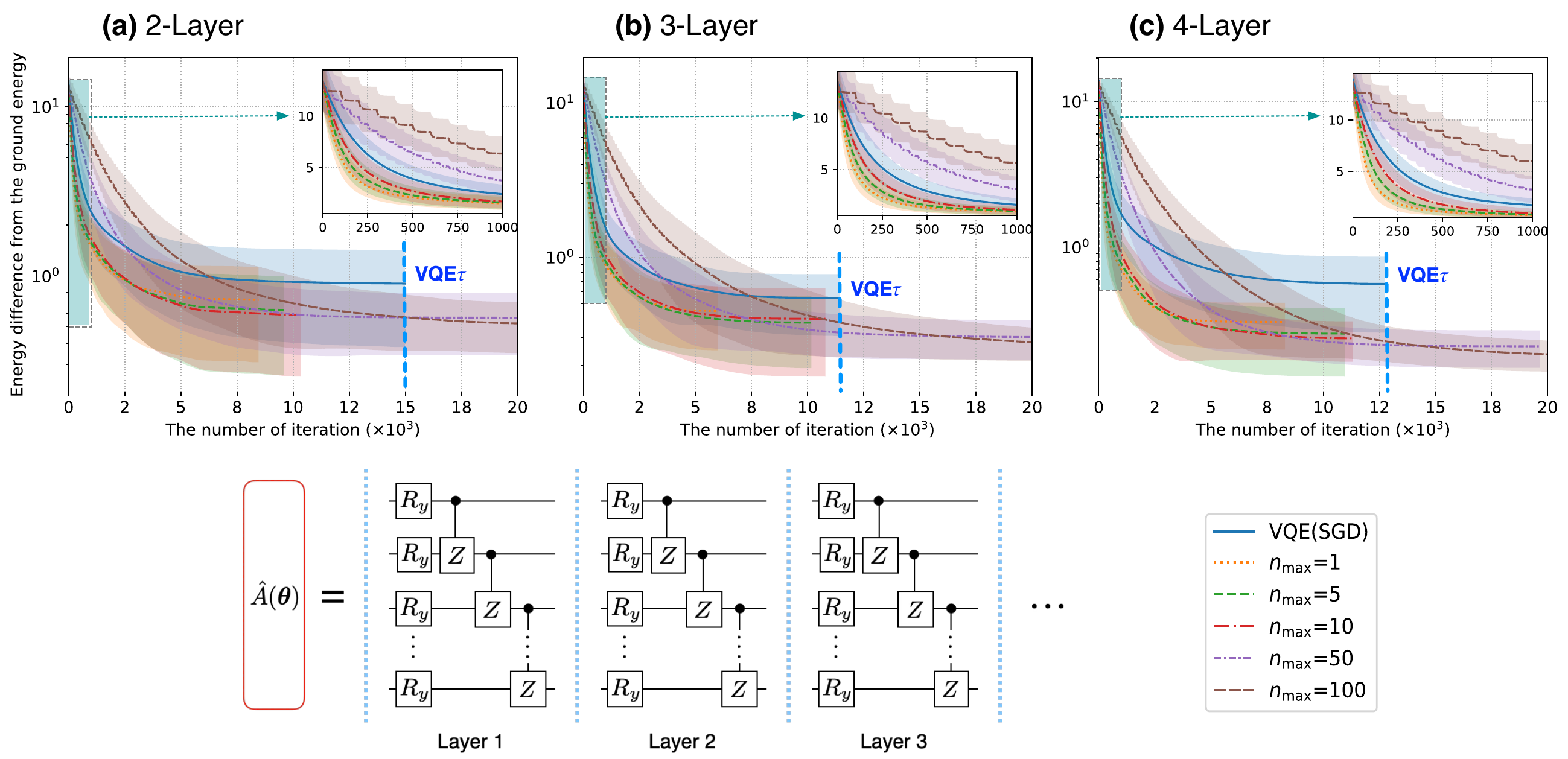}
\caption{LTFIM ($N=10$ and $(g, h)=(1, 1)$) with the hardware‑efficient ansatz: (a) $2$, (b) $3$ and (c) $4$ layers. Mean energy error (solid line) $\pm$ s.d. (shaded) over $100$ random seeds. The colors denote $n_{\max} \in \{ 100, 50, 10, 5, 1 \}$. For reference, VQE (same ansatz, optimizer, and halting rule) is overlaid. Across the depths, QAAE reaches lower errors and exhibits less variance than VQE; small $n_{\max}$ accelerates early progress but can plateau earlier, quantifying a practical speed-accuracy trade‑off.}
\label{fig:Ising}
\end{figure*}

{\em $10$‑qubit regime and comparison to VQE.}---We next probe a clearly nonintegrable LTFIM setting with $N=10$ and $(g, h)=(1, 1)$. We emulate a normalized Hamiltonian $\tfrac{1}{2}\bigl(\tfrac{1}{20}\hat{H}_{\mathrm{LTFIM}} + \hat{\openone} \bigr)$ and adopt a hardware‑efficient ansatz built from layers of $\hat{R}_y$ rotations and entangling CZ gates ($2$, $3$, and $4$ layers). For each setting, we average $100$ random initializations. The state-learning per round is budget‑limited to $n_{\max} \in \{100, 50, 10, 5, 1 \}$ learning steps on the distance‑based objective $\mathcal{F}(D)$ [Eq.~(\ref{eq:obj_fun})]. Figure~\ref{fig:Ising} plots the mean energy error vs. QAAE round and, for a fair baseline, a conventional VQE that uses the same ansatz, optimizer, and stopping rule.

The results are as follows (see Fig.~\ref{fig:Ising}): (i) Early‑stage speed: Smaller $n_{\max}$ yields faster initial error reduction because repeated amplify--learn cycles leverage the coherent overlap gains even with partially learned parameters. (ii) Final accuracy and robustness: With sufficient learning budget, QAAE attains consistently lower final errors than VQE and with smaller dispersion across random seeds, whereas VQE often stalls at a higher error floor even after $\sim 10^4$ trial states. (iii) Resource fairness: Since both methods share the same ansatz and optimizer, the gap reflects algorithmic differences. That is, QAAE learns a fixed pure target each round and does not optimizes $\bra{\Psi}\hat{H}\ket{\Psi}$; VQE, in contrast, navigates a rugged energy landscape with gradient estimates that can be fragile in depth and noise. Collectively, these results show that QAAE can surpass the gradient‑gradient-based VQE in accuracy and reliability for a nonintegrable spin model while remaining the hardware‑conscious.

\section{Discussion}\label{Sec:5}

In this work, we have developed and introduced the quantum amplitude‑amplification eigensolver (QAAE), a ground‑state (or energy) extraction framework that advances a trial state by coherently amplifying its ground‑state component and then learning the resulting pure target at each round. Note that this is not an energy‑landscape search but an amplify--learn pipeline: a state‑independent overlap gain supplied by coherent amplitude amplification, followed by a state‑learning step that re‑encodes the amplified target into the chosen ansatz. Our analysis established a monotone improvement of ground‑state overlap under standard assumptions (normalized Hamiltonian, nondegenerate ground-state, a faithful learning) and provided per‑round depth bounds that scale with the ansatz and short‑time simulation cost, ensuring hardware compatibility. We validated the mechanism on IBMQ hardware (two‑level system and a two‑qubit Ising model with an ancilla), and we benchmarked QAAE on $\mathrm{H}_2$, LiH, and a $10$‑qubit LTFIM, where it surpassed gradient‑descent VQE in accuracy and reliability while remaining depth‑conscious. These results position QAAE as a principled, non‑variational, and landscape‑free route to the ground‑state estimation for near‑term quantum simulation.

Understanding of QAAE's convergence behavior clarifies how its principle departs from the conventional variational approaches. The variational eigensolvers minimize $\bra{\Psi}\hat{H}\ket{\Psi}$ over a parameterized search space; progress is governed by the Hamiltonian's energy landscape and the ansatz trainability of the ansatz. QAAE proceeds differently: the target is a fixed pure state produced by amplitude amplification, and learning amounts to a unitary re‑encoding of that target within the ansatz family. Because the learning step targets a pure state, one may adopt distance‑based objectives augmented with barrier regularization, avoiding pathological regions while keeping the target a well‑conditioned minimizer. Thus, QAAE dispenses with energy or gradient estimators and does not traverse rugged energy landscapes, thereby reducing exposure to local minima that arise in energy‑driven searches. The induced dynamics---monotone growth of ground‑state overlap interleaved with the learning and controlled re‑encoding---are qualitatively distinct (cf.Fig.~\ref{fig:conv_behavior}), and are formalized by our ground‑overlap gain proposition together with the stability bounds for the learning step.

Per‑round circuit depth decomposes into preparation of the trial state ($\mathcal{D}_A$), a reflection about that state ($\mathcal{D}_R$), and a controlled short‑time evolution under $\hat{H}$ ($\mathcal{D}_U$). The corresponding bound scales as $5\mathcal{D}_A + 2(\mathcal{D}_R + \mathcal{D}_U)$ with $\mathcal{D}_R \le 2\mathcal{D}_A + \mathcal{O}(q)$ and $\mathcal{D}_U=\mathcal{O}\bigl(\mathrm{poly}(q) \epsilon_H^{-1/2\kappa}\bigr)$, so the implementability is dictated chiefly by the ansatz architecture and the simulation primitive for $e^{-i\omega\hat{H}}$ (Trotter/QSP/LCU). In practice, the chemistry‑inspired (e.g., UCCSD) and hardware‑efficient ansatz can both be embedded in QAAE; the former provides chemically structured expressivity, while the latter offers shallow and device‑tailored circuits for NISQ experiments. This flexibility underpins our polynomial‑depth implementations across the studied systems.

The QAAE's guarantees rest on (i) normalized $\hat{H}$ with a nondegenerate ground level, (ii) accurate short‑time simulation $e^{-i\omega\hat{H}}$, and (iii) faithful learning/re‑encoding in the chosen ansatz. The expressivity limits can slow or stall the learning if the amplified target exits the ansatz manifold; our experiments quantify this trade‑off and show that scheduling the per‑round learning budget $n_{\max}$ (greedy in early rounds, precise in later rounds) provides practical control. As devices improve, error‑mitigation and/or noise‑aware objective shaping (e.g., normalization of the measured Pauli vectors) can further bolster robustness, as already observed in our hardware runs.

Looking ahead, QAAE invites several focused extensions. First, the learning step for pure‑state targets can be refined via geometry‑aware, distance‑based objectives with barrier terms, adaptive per‑round learning budgets, and noise‑aware normalization of the measured observables. Second, we can exploit a partial structure of $\hat{H}$---symmetries, locality, or approximate spectra---to restrict the re‑encoding manifold and accelerate the convergence. Lastly, it would be possible to curb the per-round depth by improving short‑time simulation (e.g., higher‑order Trotter, QSP/LCU with lightweight block encodings). We expect these directions to broaden QAAE's reach across chemistry, materials, and many‑body physics on increasingly capable devices.

\section*{Acknowledgement}

JB thanks Dr.~Antonio Mandarino. JB and MSK thanks Jinzhao Sun for discussions. This work was supported by the Ministry of Science, ICT and Future Planning (MSIP) by the National Research Foundation of Korea (RS-2024-00432214, RS-2025-03532992, and RS-2023-NR119931), the Institute of Information and Communications Technology Planning and Evaluation grant funded by the Korean government (RS-2019-II190003, ``Research and Development of Core Technologies for Programming, Running, Implementing and Validating of Fault-Tolerant Quantum Computing System''), the Korean ARPA-H Project through the Korea Health Industry Development Institute (KHIDI), funded by the Ministry of Health \& Welfare, Republic of Korea (RS-2025-25456722), and the Ministry of Trade, Industry, and Energy (MOTIE), Korea, under the project ``Industrial Technology Infrastructure Program'' (RS-2024-00466693). MSK acknowledges the  UK EPSRC EP/W032643/1 and EP/Z53318X/1, the KIST through Open Innovation fund and the National Research Foundation of Korea (NRF) grant funded by the Korean Government (MSIT) (RS-2024-00413957). The entirety of SL's contribution to the present work was completed while affiliated with his former institution, Electronics and Telecommunications Research Institute (ETRI). We acknowledge the Yonsei University Quantum Computing Project Group for providing support and access to the Quantum System One (Eagle Processor), which is operated at Yonsei University.

\appendix

\bibliography{mybibfile_QAAE}

\end{document}